\documentclass[reprint,
superscriptaddress,
amsmath,amssymb,aps,showkeys,showpacs,
twoside,final,secnumarabic,
nofootinbib]{revtex4-2}

\usepackage[paperwidth=205mm,paperheight=290mm,top=17mm,bottom=25mm,
inner=17mm,outer=17mm,
twoside]{geometry}

\usepackage{cmap} 
\usepackage[T1,T2A]{fontenc}
\usepackage[utf8]{inputenc}
\usepackage[russian,english]{babel}
\usepackage{color}
\usepackage{graphicx}
\usepackage{dcolumn}
\usepackage{bm} 
\usepackage[unicode=true,colorlinks=true,linkcolor=magenta, urlcolor=blue, citecolor = blue,breaklinks]{hyperref}
\usepackage{multirow}
\usepackage{url}
\usepackage{breakurl}
\DeclareGraphicsExtensions{.eps}

\newcount\issue
\newcount\Vol
\newcount\numb
\usepackage{fancyhdr} 
\def\Vol{\textbf{78}}
\def\numb{x}
\begin{document}

\title{
Cosmic Inflation, Dark Energy and Gravitational Waves} 

\def\addressa{Consortium for Fundamental Physics, Physics Department,
Lancaster University, Lancaster LA1 4YB, UK}

\author{\firstname{K.}~\surname{Dimopoulos}}
\email[E-mail: ]{k.dimopoulos1@lancaster.ac.uk}
\affiliation{\addressa}

\received{xx.xx.2023}
\revised{xx.xx.2023}
\accepted{xx.xx.2023}

\begin{abstract}
  We briefly discuss cosmic inflation, which is the dominant paradigm
  for the generation of the large scale structure in the Universe and also
  for arranging for the initial conditions of the hot Big Bang. We then present
  quintessential inflation, which also accounts of the observed dark energy.
  We discuss how quintessential inflation can be successfully modelled in
  modified gravity in the Palatini formalism. Finally, we focus on the
  generation of primordial gravitational waves by inflation and how their
  spectrum can be
  enhanced when the early Universe goes through periods of stiff equation of
  state. This results in gravitational waves with a characteristic spectrum,
  which may well be observed in the near future, providing insights for the
  background theory.  
\end{abstract}

\pacs{Suggested PACS}\par
\keywords{Suggested keywords   \\[5pt]}

\maketitle
\thispagestyle{fancy}


\section{
Cosmic Inflation}\label{cosmicinf}

The history of the Universe requires special initial conditions, which are
arranged by cosmic inflation \cite{guth,staro}.
In a nutshell, cosmic inflation can be defined as
a period of accelerated expansion in the Early Universe
\cite{mybook,davidsbook}.
Inflation produces
a Universe which is large, uniform and spatially flat according to observations.
Typically, inflation is realised via the inflationary paradigm, which states
that the Universe inflates when dominated by the potential energy density of
a scalar field, called the inflaton field.

\begin{figure}[h]
\vspace{-3cm}
\centering
\includegraphics[width=8cm]{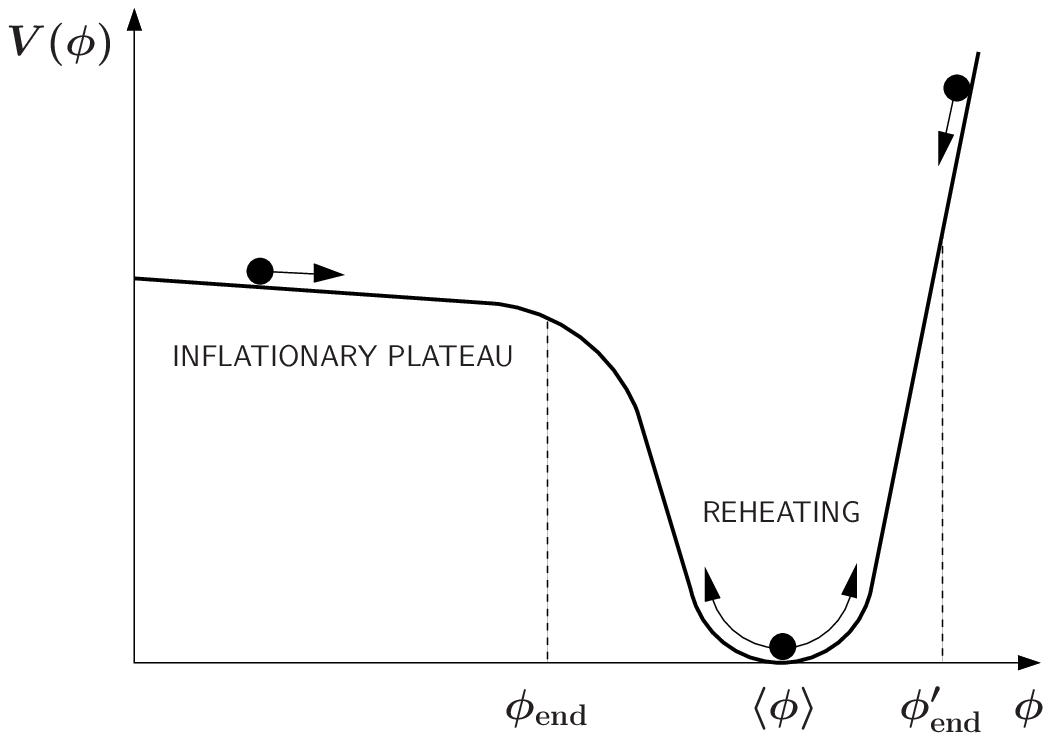}
\vspace{-3cm}
\caption{\label{infpot}
  Sketch of the typical inflationary potential. Due to the form of the
  equation of motion \eqref{KG}, we can envisage the system as a ball rolling
  down a flat region of the potential, called the inflationary plateau. At some
  critical value of the inflaton field $\phi_{\rm end}$, the potential becomes
  steep and curved such that inflation is terminated. Afterwards, the field
  oscillates around its vacuum expectation value $\langle\phi\rangle$. The
  figure also depicts the possibility that the inflaton field slow-rolls down a
  steep potential under excessive friction (and inflation ends when this
  friction is not enough for inflation at $\phi'_{\rm end}$) but this possibility
  is not favoured by the observations.}
\end{figure}

The Klein-Gordon equation of motion of a homogeneous scalar field $\phi$ is
\begin{equation}
\ddot\phi+3H\dot\phi+V'(\phi)=0\,,
\label{KG}
\end{equation}
where $H$ is the rate of the Universe expansion (Hubble parameter), the
dot denotes derivative with respect to the cosmic time $t$ and the prime
denotes derivative with respect to the field: $'\equiv\partial/\partial\phi$.
The above is of the same form as the equation of motion of a ball sliding down
a potential under friction determined by $H$ (see Fig.~\ref{infpot}).
Potential domination,
therefore, suggests that the kinetic energy density is subdominant to the
potential energy density $V$, and the field slowly rolls (slowly varies in field
space) down a potential plateau, called the inflationary plateau. Inflation ends
at a characteristic value $\phi_{\rm end}$ when the potential becomes steep and
curved. After the end of inflation, the inflaton field oscillates around its
vacuum expectation value (VEV). These coherent oscillations amount to inflaton
particles, which decay into the primordial plasma, through a process called
reheating.

\begin{figure}[h]
\vspace{-4.5cm}
\centering
\mbox{\hspace{-1.3cm}\includegraphics[width=10.5cm]{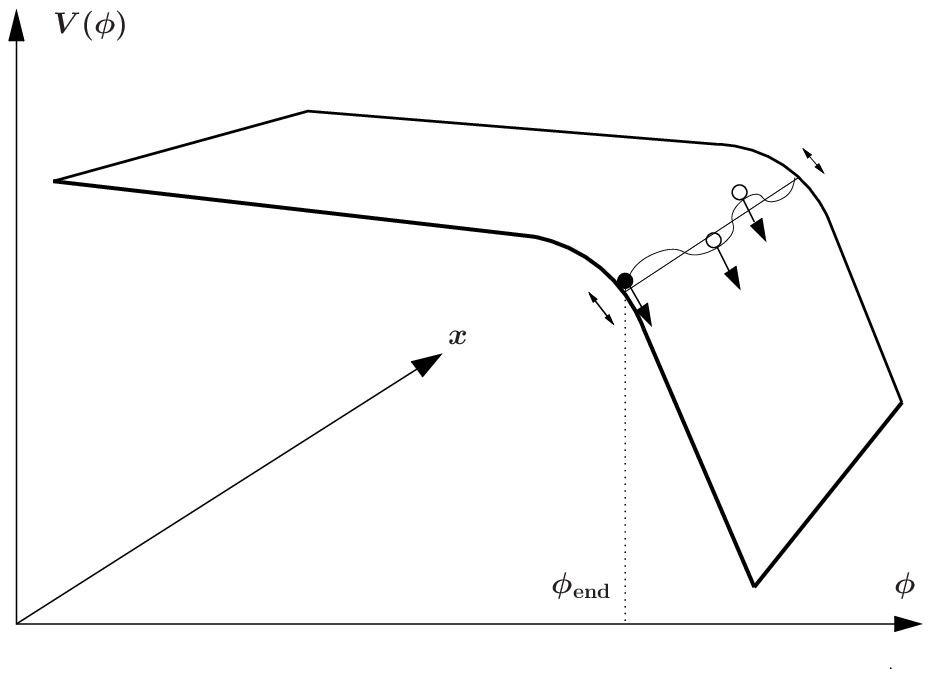}}
\vspace{-4.5cm}
\caption{\label{infend}
  Sketch of the edge of the inflationary plateau, where inflation is terminated
  at the critical value $\phi_{\rm end}$. The spatial direction $x$ is also
  depicted (assumed one-dimensional for illustrative purposes). Perturbations
  of the inflaton field $\phi$ imply that, while it is rolling down its
  potential, it reaches the critical value $\phi_{\rm end}$ at different times
  at different locations (values of $x$). This means that inflation continues a
  little bit more in some locations than in others, which leads to the
  generation of the primordial density perturbations.}
\end{figure}

Inflation however, should not make the Universe perfectly uniform, because in
order for galaxies to form, initial perturbations in the density of the Universe
are needed. Indeed, inflation makes the Universe largely uniform but also
introduces minor deviations from uniformity which give rise to the
Primordial Density Perturbations (PDPs), which in turn become the seeds for the
formation of structures such as galaxies \cite{davidsbook}.
Inflation does this through the
particle productions process which roughly operates as follows:

Accelerated expansion of space is superluminal. This superluminal expansion
during inflation amplifies the quantum fluctuations of the inflaton field,
which become classical perturbations of the field through quantum decoherence.
Consequently, inflation continues a little bit more in some locations than in
others. Thus, at the end of inflation space expands in a different way in
neighbouring locations, which introduces the PDPs (see Fig.~\ref{infend}).

\begin{figure}[h]
\centering
\includegraphics[width=8cm]{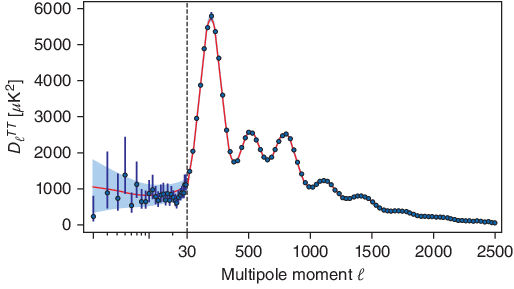}
\caption{\label{Planck2015}
  CMB temperature perturbation spectrum. The graph is a plot of
  \mbox{${\cal D}_\ell=\ell(\ell+1)C_\ell/2\pi$}, where $C_\ell$ is the CMB
  temperature anisotropy, as a function of the multipole
  moment $\ell$ of spherical harmonics. For \mbox{$\ell<30$} the
  scaling of the horizontal axis changes. The shaded area
  depicts the region under the influence of cosmic variance, which suggests that
  deviations from the theoretical curve (depicted by the solid line) within the
  shaded area are due to poor statistics and do not have a physical meaning. The
  effect of cosmic variance diminishes (virtually disappears) for large $\ell$.
  For \mbox{$\ell\geq 30$} we have the clear depiction of seven acoustic peaks.
  The line between the binned data points is not the line which connects the
  dots; it is the theoretical line predicted by inflation. The agreement with
  the data is impressive. Figure taken from Ref.~\cite{planck}.}
\end{figure}

The PDP reflects itself onto the Cosmic Microwave Background radiation (CMB)
through the Sachs-Wolfe effect \cite{SW}.
Indeed, precise CMB observations have revealed
the existence of the PDP at the level of $\sim 10^{-5}$, with the
characteristics suggested by inflation (acoustic peaks). The agreement with
the observations (see Fig.~\ref{Planck2015}) is spectacular \cite{planck}.

The PDPs are predominately adiabatic, Gaussian and scale invariant
\cite{davidsbook}. Adiabaticity
suggests that they are the product of a single degree of freedom, such as the
inflaton field. Gaussianity reflects the randomness of the original quantum
fluctuations. Approximate scale invariance suggests that inflation is of
quasi-de Sitter type, when the density $\rho$ is roughly constant during
inflation. The barotropic (equation of state) parameter of a homogeneous scalar
field is
\begin{equation}
  w\equiv\frac{p}{\rho}=
  \frac{\frac12\dot\phi^2-V(\phi)}{\frac12\dot\phi^2+V(\phi)}\,,
\label{w}
\end{equation}
where $p$ is the pressure. According to the inflationary paradigm, the kinetic
energy density $\rho_{\rm kin}$ is subdominant to the potential during inflation,
which means \mbox{$\rho_{\rm kin}\equiv\frac12\dot\phi^2\ll V$}. As a result,
during quasi-de Sitter inflation we have \mbox{$w\approx -1$}. 

\begin{figure}[h]
\centering
\mbox{\hspace{-.5cm}\includegraphics[width=8.5cm]{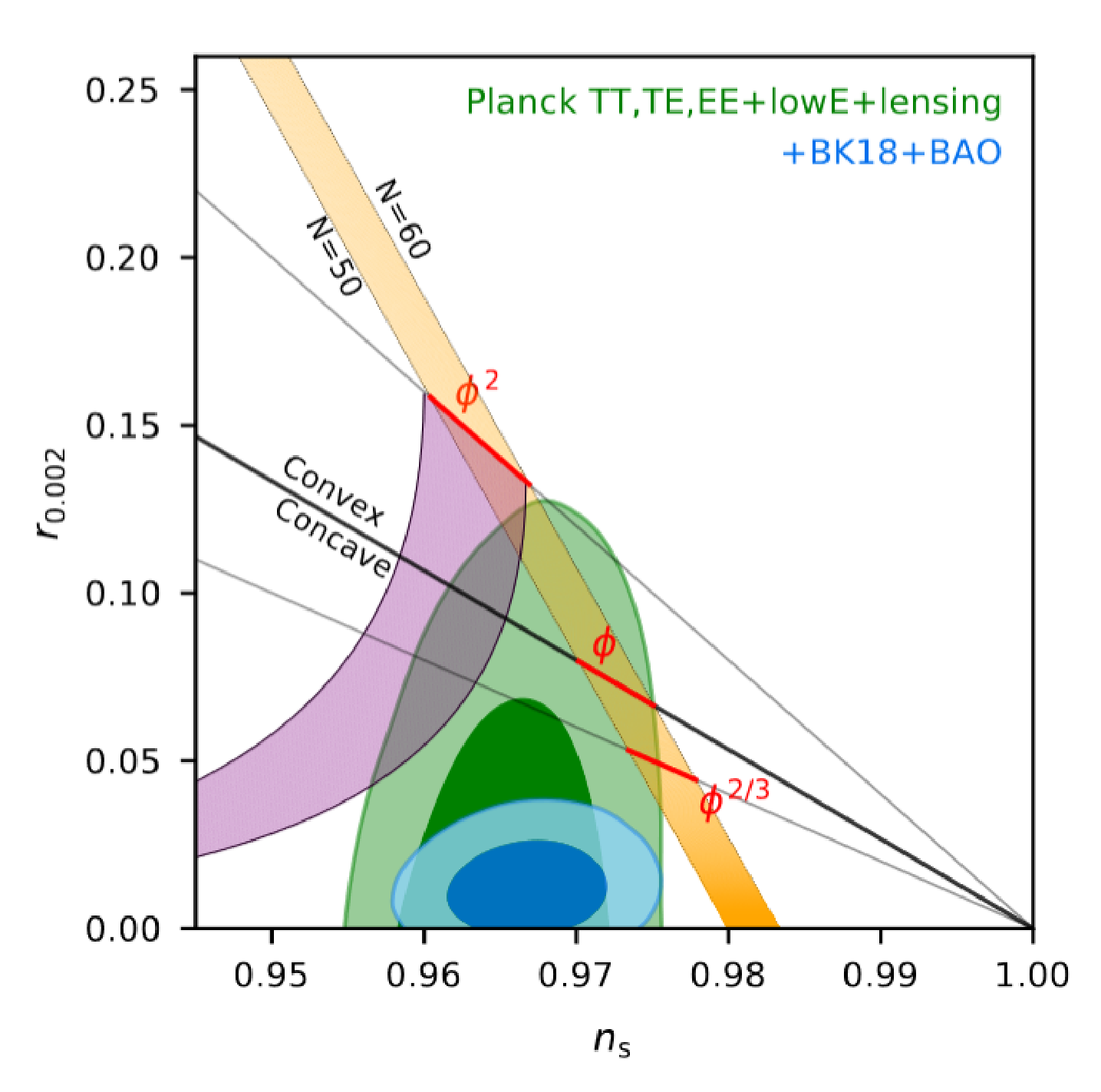}}
\caption{\label{KeckPlot}
  Observations of BICEP/Keck Collaboration
  plot the tensor to scalar ratio $r$ with respect to
  the scalar spectral index $n_s$. The allowed parameter space is shown in
  blue (contours depict the 95\% c.l. [light] and 67\% c.l. [dark]).
  Observations exclude families of otherwise motivated inflation models.
  For example, the clearly excluded purple band and orange band correspond to
  natural and chaotic monomial inflation models respectively. The figure was
  taken from Ref.~\cite{keck}.}
\end{figure}

There is a lot of emphasis put on the PDP generated by inflation, because the
perturbations generated can discriminate between different inflation models.
In particular, two observables are of prime interest: the scalar spectral index
$n_s$ and the tensor-to-scalar ratio $r$. For the scalar perturbations, the
spectrum can be written as \mbox{${\cal P}_\zeta(k)\propto k^{n_s(k)-1}$}, where
$k$ is the momentum scale. A perfectly scale invariant spectrum would
correspond to \mbox{$n_s=1$}, in which case all $k$-dependence of
${\cal P}_\zeta$ would disappear. Indeed, observations suggest that, for
negligible tensor perturbations, the spectral index is very close to unity
\mbox{$n_s=0.965\pm0.004$} \cite{planck}. Crucially, the spectral index is not
exactly equal
to unity because the end of inflation is near and the inflationary plateau is
not exactly flat. Therefore, a slightly red spectrum is expected, exactly as
observed. The tensor-to-scalar ratio is
\mbox{$r\equiv{\cal P}_\zeta/{\cal P}_t$}, where ${\cal P}_t$ is the spectrum of
the tensor perturbations (primordial gravitational waves). Observations suggests
that there is a stringent upper limit \mbox{$r<0.036$} \cite{keck},
which can be even
tighter under certain conditions. Precision observations of $n_s$ and $r$ have
already resulted in the exclusion of many, otherwise well motivated inflation
models (see Fig.~\ref{KeckPlot}).

\subsection{\boldmath $R^2$ inflation}\label{R2}

The seminal paper of Alexei A.~Starobinsky, which appeared in 1980, even before
the name `cosmic inflation' was coined, introduced the very first and the most
successful to date inflation model \cite{staro}. It is a modified gravity
theory with a simple Lagrangian density
\begin{equation}
{\cal L}=\frac12 m_P^2 R+\alpha R^2\,,
\label{R2}
\end{equation}
where $m_P$ is the reduced Planck mass and $R$ is the scalar curvature (Ricci
scalar). The first term in the above corresponds to the standard
Einstein-Hilbert action. However, the second term, whose importance is
parametrised by the non-perturbative coefficient $\alpha$, corresponds to
modified gravity. This quadratic gravity term introduces an additional degree of
freedom. The latter can be flushed out if we switch from the modified gravity
frame (called the Jordan frame) to the frame of Einstein gravity (called the
Einstein frame) through a conformal transformation of the spacetime metric,
of the form \mbox{$g_{\mu\nu}\rightarrow\Omega^2 g_{\mu\nu}$}, where the conformal
factor for this theory is \mbox{$\Omega^2=1+\frac{4\alpha}{m_P^2}R$}. In the
Einstein frame the Lagrangian density becomes
\begin{equation}
{\cal L}=\frac12 m_P^2 R+\frac12(\partial\phi)^2-V(\phi)\,,
\label{EinsteinL}
\end{equation}
where $R$ is now calculated using the new metric and, apart from the
Einstein-Hilbert term, $\cal L$ features a minimally coupled (to gravity), 
canonically normalised scalar field
$\phi$, where \mbox{$(\partial\phi)^2\equiv-\partial_\mu\phi\,\partial^\mu\phi$}.
This scalar field corresponds to the new degree of freedom introduced by the
original quadratic gravity term, and it is called the scalaron field with
\mbox{$\Omega^2=\exp\left(\sqrt{\frac23}\,\phi/m_P\right)$}.

\begin{figure}[h]
\vspace{-4cm}
\centering
\includegraphics[width=8cm]{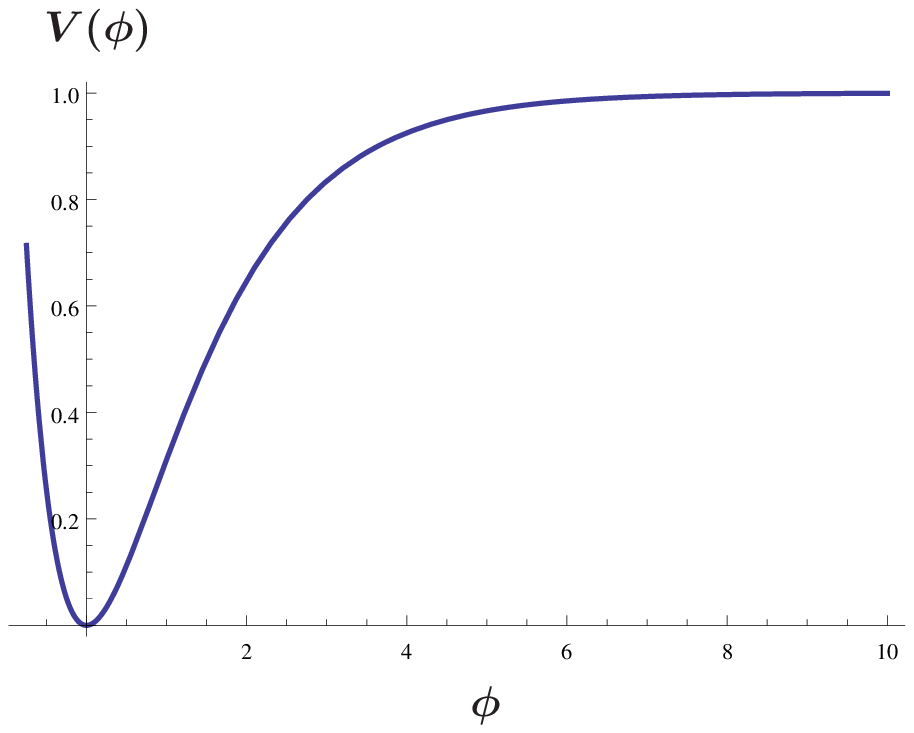}
\vspace{-2cm}
\caption{\label{staro}
  Form of the Starobinsky potential, which applies to both the cases of
  $R^2$-inflation and Higgs inflation. The plot clearly depicts the inflationary
  plateau at large values of the inflaton field. At small values, the potential
  becomes steep and curved and the inflaton oscillates around its VEV, which is
  zero in this case. Fiducial units are used.}
\end{figure}

For this theory, in the Einstein frame, the scalar potential is
\begin{equation}
V(\phi)=\frac{m_P^4}{16\alpha}\left(1-e^{-\sqrt{\frac23}\,\phi/m_P}\right)^2\,.
\label{Vstaro}
\end{equation}
The form of the above potential is shown in Fig.~\ref{staro}. The model predicts
\begin{equation}
n_s=1-\frac{2}{N}\quad{\rm and}\quad r=\frac{12}{N^2}\,,
\label{nsrstaro}
\end{equation}
where $N$ is the number of e-folds (exponential expansions) of remaining
inflation when the cosmological scales exit the horizon during inflation (they
are pushed out by the superluminal expansion). Typically, \mbox{$N=50-60$}
depending on the details of reheating (prompt reheating results in $N=60$).
The above suggest that the Starobinsky inflation predictions are in the sweet
spot of the observations. For the amplitude of the PDP to match the CMB
observations we require \mbox{$\alpha=5.5225\times 10^8$}.

\subsection{Higgs inflation}\label{Higgsinf}

Another seminal inflation model was put forward by Fedor~L.Bezrukov and
Mikhail Shaposhnikov in 2008 \cite{higgsinf}.
The model considers the electroweak Higgs field
$h$ (first observed in CERN in 2012) as the inflaton. It is another modified
gravity theory with Lagrangian density
\begin{equation}
{\cal L}=\frac12 m_P^2 R+\frac12\xi Rh^2+\frac12(\partial h)^2-U(h)\,,
\label{Higgs}
\end{equation}
where $\xi$ parametrises the strength of the non-minimal coupling to gravity
of the Higgs field, which is naturally introduced by quantum corrections in
curved spacetime. The potential is \mbox{$U(h)=\frac14\lambda_h(h^2-v^2)^2$},
where \mbox{$v=246\,$GeV} is the Higgs VEV and $\lambda_h$ is its self-coupling,
which at low energies is \mbox{$\lambda_h=0.129$}.

We switch to Einstein gravity using the conformal transformation
\mbox{$\Omega^2=1+\frac{\xi h^2}{m_P^2}$}. In the Einstein frame the Lagrangian
density is given by Eq.~(\ref{EinsteinL}) with potential
\begin{equation}
V(\phi)=\frac{\lambda_h m_P^4}{4\xi^2}\left(1-e^{-\sqrt{\frac23}\,\phi/m_P}\right)^2\,.
\label{VHiggs}
\end{equation}
The above is of exactly the same form as the potential in Eq.~\eqref{Vstaro} if
we identify \mbox{$\xi^2=4\alpha\lambda_h$}. Thus, the inflationary predictions
are equally successful. In fact, because the branching rations of the Higgs
decays into the standard model particles are known, reheating is unambiguously
defined and \mbox{$N\simeq 57$}. As a result, Eq.~\eqref{nsrstaro} suggests
\mbox{$n_s=0.965$} and \mbox{$r=0.0037$}, which are in excellent agreement with
observations. To have the correct amplitude of the PDP we require
\mbox{$\xi\simeq 47000/\sqrt{\lambda_h}$}.

\section{Dark Energy}

Observations suggest that the late Universe is also undergoing accelerated
expansion. This is attributed to an exotic substance called dark energy, which
makes up almost 70\% of the Universe content at present \cite{DE}.
Dark energy could correspond to a positive cosmological constant $\Lambda$ but
this requires phenomenal fine-tuning (of order $10^{-120}$) which has been called
``the worse fine-tuning in physics'' (Lawrence Krauss). As a result, other
proposals have been put forward. A prominent such suggestion is quintessence
\cite{Q},
which amounts to the fifth element after baryonic (normal) matter, dark matter,
photons (mainly CMB) and neutrinos. Quintessence is a scalar field like the
inflaton, typically slow-rolling down a runaway potential; a feature called the
quintessential tail. Thus, the dark energy observations suggest that the
Universe is undergoing a late inflation period, driven by quintessence. This
late inflation is also quasi-de Sitter, with barotropic parameter
\mbox{$-1\leq w\leq -0.95$} \cite{planck}.

Being a dynamical degree of freedom, quintessence introduces an additional
requirement (compared with just $\Lambda$), that of its initial conditions.
Indeed, in thawing quintessence the field is initially frozen only to unfreeze
near the present time to become dominant. But the initial (frozen) value of the
field must be such that the potential energy density of the field today is
comparable ($\sim$ 70\%) to the density of the Universe at present. This is
called the coincidence requirement \cite{DE}.

\section{Quintessential Inflation}

A promising way to overcome the coincidence requirement is the idea of
quintessential inflation, proposed first by P.~James E.~Peebles and
Alexander Vilenkin in 1999 \cite{PeebVil}. In a nutshell, quintessential
inflation is the identification of quintessence with the inflaton field.

Quintessential inflation is a natural idea, as quintessence and the inflaton are
both scalar fields. It is also economic because one employs only a single
degree of freedom. Furthermore, it allows to treat the physics of inflation
in the early Universe and dark energy in the late Universe in a common
theoretical framework, valid over a vast range of energies. Quintessential
inflation has to satisfy the observations both cosmic inflation and of dark
energy, which is very difficult but not impossible. Finally, the initial
conditions of quintessence (the coincidence requirement) are fixed by the
inflationary attractor.

\begin{figure}[h]
\vspace{-3.5cm}
\centering
\includegraphics[width=8cm]{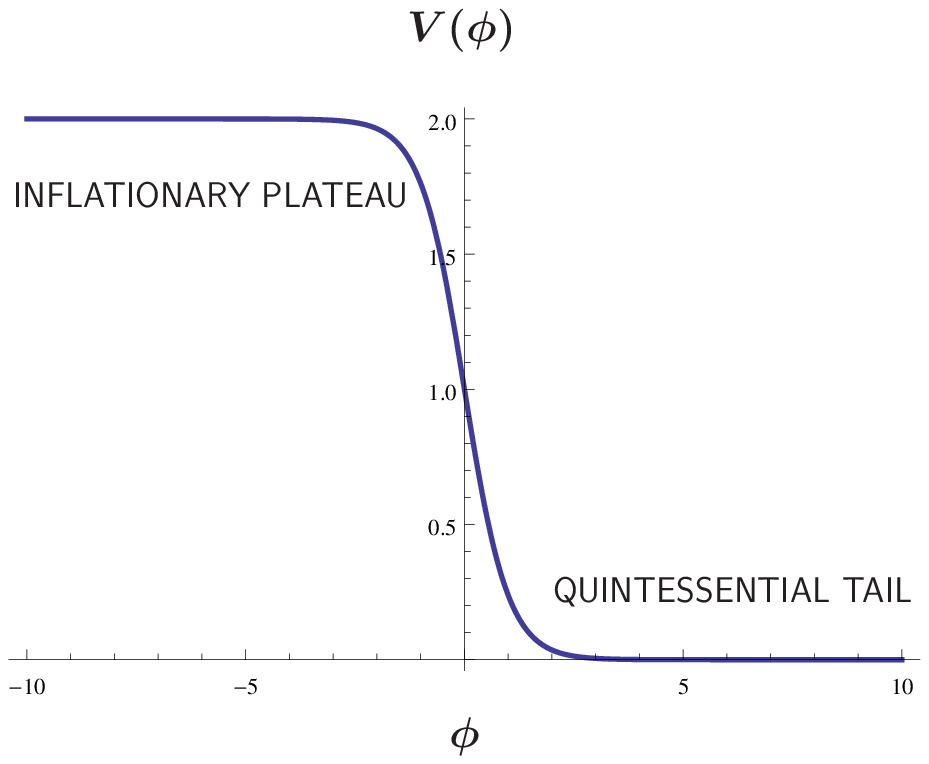}
\vspace{-3cm}
\caption{\label{quinfpot}
  The typical potential of quintessential inflation. It depicts a runaway
  direction with VEV displaced at infinity. There are two flat regions, the
  inflationary plateau and the quintessential tail, chosen (without loss of
  generality) to be at small and large values of the scalar field.}
\end{figure}

The potential of quintessential inflation models typically features two flat
regions, the inflationary plateau and the quintessential tail \cite{mybook}
(see Fig.~\ref{quinfpot}). In
contrast to the traditional inflationary paradigm, the inflaton field does not
decay after the end of inflation, because it has to survive until the present
time and become quintessence. Consequently, reheating has to occur without the
decay of the inflaton field. Fortunately, there are many motivated mechanisms
that achieve this.

\subsection{Quintessential inflation in Palatini gravity}\label{Palatini}

A promising framework for the construction of quintessential inflation models
is modified gravity on the Palatini formalism \cite{samuel}.
The Palatini formalism considers
both the spacetime metric and the connection to be independent degrees of
freedom, in contrast to the metric formalism, where the connection is the
Levi-Civita. With the Einstein-Hilbert action, both formalisms result in
general relativity. However, modified gravity actions result in different
theories when assuming the metric or the Palatini formalism. Here we will argue
that the Palatini formalism gives rise naturally to a potential suitable for
quintessential inflation.

We consider the Lagrangian density in the Jordan frame
\begin{equation}
{\cal L}=\frac12 m_P^2R+\alpha R^2+\frac12(\partial\varphi)^2-V(\varphi)\,,
\label{PalatiniJordan}
\end{equation}
where we considered a quadratic gravity term, as in Starobinsky inflation
discussed in Sec.~\ref{R2}, and
we have explicitly introduced a canonically normalised scalar field $\varphi$.
We did this because, in the Palatini formalism, the quadratic gravity term $R^2$
does not introduce an additional degree of freedom (there is no scalaron). 

Using a suitable conformal transformation, we switch to the Einstein frame,
where the Lagrangian density is
\begin{equation}
  {\cal L}=\frac12 m_P^2R+
  \frac{\frac12(\partial\varphi)^2}{1+16\alpha\frac{V(\varphi)}{m_P^4}}-
  \frac{V(\varphi)}{1+16\alpha\frac{V(\varphi)}{m_P^4}}\,.
\label{PalatiniEinstein}
\end{equation}
We see that in the Einstein frame the scalar field has a non-canonical kinetic
term. Notice however, that, when $V$ is very large, the unity in the denominator
is negligible and the potential density approaches the constant value:
\mbox{$m_P^4/16\alpha$}, This corresponds to the inflationary plateau, as the
potential energy density does not depend of the value of the inflaton field,
i.e. it is the same for a range of $\varphi$ values.

In the opposite limit, when $V$ is very small then the denominator becomes unity
and the field becomes canonically normalised. Thus, we only need a suitable
runaway potential to arrange for the quintessential tail, while the Palatini
setup generates the inflationary plateau by ``flattening'' the potential at high
energies.

We apply this finding in the following setup. Now the Lagrangian density
in the Jordan frame is \cite{sameemeli}
\begin{equation}
  {\cal L}=\frac12 m_P^2R+\alpha R^2+\frac12\xi R\varphi^2+
  \frac12(\partial\varphi)^2-V(\varphi)\,,
\label{PalatiniJordan+}
\end{equation}
where we also consider a non-minimal coupling of $\varphi$ to gravity, which is
expected naturally by quantum corrections in curved spacetime, as we mention in
Sec.~\ref{Higgsinf}.

After performing a conformal transformation we bring the theory in the Einstein
frame. We also redefine a canonically normalised inflaton field $\phi$ by
using the relation
\begin{equation}
  \left(\frac{{\rm d}\phi}{{\rm d}\varphi}\right)^2=
  \frac{1+\xi\varphi^2/m_P^2}{(1+\xi\varphi^2/m_P^2)^2+
    \frac{16\alpha}{m_P^4}V(\varphi)}\,.
  \label{phivarphi}
\end{equation}
Then, the Lagrangian density becomes
\begin{equation}
{\cal L}=\frac12 m_P^2 R+\frac12(\partial\phi)^2-U(\phi)\,,
\label{EinsteinL+}
\end{equation}
where the potential is
\begin{equation}
  U(\phi)=\frac{V(\varphi(\phi))}{[1+\xi\varphi^2(\phi)/m_P^2]^2+
   \frac{16\alpha}{m_P^4}V(\varphi(\phi))}\,.
\label{U}
\end{equation}
We see that, again, if $V$ is very large, we approach the inflationary plateau
with \mbox{$U\simeq m_P^4/16\alpha$}. We only need a runaway potential. We
choose a simple exponential potential \cite{sameemeli},
which is amply motivated by fundamental theory:
\begin{equation}
V(\varphi)=V_0\,e^{-\kappa\varphi/m_P}\,,
\label{V}
\end{equation}
where $\kappa$ is the strength of the exponential. As we approach the
quintessential limit 
the potential of the quintessential tail approaches the form \cite{sameemeli}
\begin{equation}
  U(\phi)\simeq V_0\frac{\exp
\left[-\frac{\kappa}{\sqrt\xi}\sinh\left(\sqrt\xi\frac{\phi}{m_P}\right)\right]}{\cosh^4\left(\sqrt\xi\frac{\phi}{m_P}\right)}\,,
\label{UDE}
\end{equation}
where we used the fact that, at late times, the quadratic gravity term is
negligible, which means that Eq.~\eqref{phivarphi} can be solved exactly to give
\mbox{$\sqrt\xi\varphi=m_P\sinh\left(\sqrt\xi\phi/m_P\right)$}. In the limit
\mbox{$\sqrt\xi\varphi\ll m_P$}, we recover the classic exponential
quintessential tail \mbox{$U(\phi)\simeq V_0\,e^{-\kappa\phi/m_P}$}.

\section{Kination}

The Universe history in quintessential inflation typically includes a period,
just after inflation, where the Universe is dominated by the kinetic energy
density of the rolling scalar field \mbox{$\rho_{\rm kin}=\frac12\dot\phi^2$}.
This period is called kination \cite{kin}
and usually follows the end of inflation until
reheating, when the radiation era of the hot Big Bang begins.

Indeed, as the inflaton rolls off the inflationary plateau, it plunges down
the potential cliff (see Fig.~\ref{quinfpot}) and becomes kinetically dominated,
i.e. \mbox{$V\ll\rho_{\rm kin}$}. Because the inflaton continues to dominate the
Universe, according to Eq.~\eqref{w}, the barotropic barotropic parameter of
the Universe is \mbox{$w\approx 1$}. The Klein-Gordon in Eq.~\eqref{KG} becomes
oblivious of the potential: \mbox{$\ddot\phi+3H\dot\phi\simeq 0$},
whose solution suggests that \mbox{$\rho=\rho_{\rm kin}\propto a^{-6}$}, where
$a(t)$ is the scale factor of the Universe. During kination the field rolls down
to the quintessential tail of its potential.

Because the energy density of radiation is diluted less efficiently with the
Universe expansion, \mbox{$\rho_r\propto a^{-4}$}, if some mechanism generates
some (initially subdominant) radiation at the end of inflation (or afterwards),
this radiation eventually catches up with the kinetic energy density of the
rolling
scalar field and the Universe becomes radiation dominated. This is the moment of
reheating \mbox{$\rho_{\rm kin}^{\rm reh}=\rho_r^{\rm reh}$}, the onset of the hot Big
Bang.

\begin{figure}[h]
\vspace{-3.5cm}
\mbox{\hspace{-1cm}\includegraphics[width=10cm]{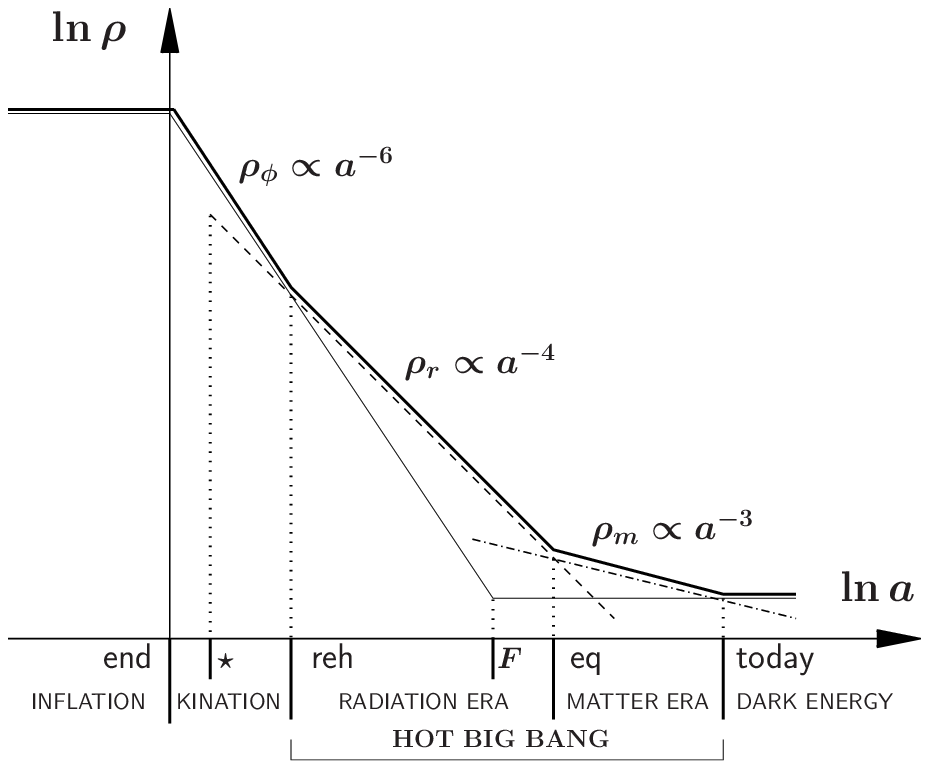}}
\vspace{-4.5cm}
\caption{\label{kin}
  Schematic log-log plot of the density of the Universe and its
  individual components with respect to the scale factor according to
  quintessential inflation. The thin solid line
  depicts the density of the scalar field, which is roughly constant during
  inflation (i.e. until inflation ends, denoted by `end') and after freezing
  (denoted by `$F$') but scales as \mbox{$\rho_\phi\propto a^{-6}$} in between,
  regardless of whether it is dominant or not. The dashed line depicts the
  density of radiation (formed at the moment denoted by `$\star$', but it is
  identified with `end' in most cases), which scales
  as \mbox{$\rho_r\propto a^{-4}$}. Radiation dominates at reheating (denoted by
  `reh'), which marks the onset of the hot Big Bang. The dot-dashed line depicts
  the density of matter, which scales as \mbox{$\rho_m\propto a^{-3}$} and
  takes over from radiation at the time of equality (denoted by `eq'). Today is
  the end of the hot Big Bang, when the density of 
  the scalar field takes over again accounting for dark energy. The thick solid
  line depicts the total density.}
\end{figure}

After the radiation era begins, it can be shown that the field's roll is
impeded and soon the field freezes in some value $\phi_F$ (see Fig.~\ref{kin}).
Because the roll of the scalar field is oblivious of the potential during
kination and afterwards until it freezes, it can be studied in a model
independent way. In particular, the total excursion of the scalar field in field
space from the end of inflation and until it freezes is \cite{mine}
\begin{equation}
  \Delta\phi_F=\sqrt{\frac23}\left(2-\frac32\ln\Omega_r^{\rm end}\right)\,m_P\;,
\label{DphiF}
\end{equation}
where \mbox{$\Omega_r^{\rm end}=(\rho_r/\rho)_{\rm end}$} is the density parameter
of radiation at the end of inflation, sometimes called reheating efficiency.%
\footnote{Here, we assume that the mechanism which generates the radiation bath
  operates at the end of inflation, as it is typically the case.}
Thus, we see that the smaller the reheating efficiency $\Omega_r^{\rm end}$ is
the largest the roll of the scalar field after inflation and until it freezes
and the longer the corresponding kination period is. The significance of this
is discussed next.

\section{Stiff Period and Gravitational Waves}

In the inflationary paradigm, inflation generates an almost scale invariant
superhorizon spectrum of tensor perturbations (gravitational waves) \cite{GW}.
This is so
only for the scales which re-enter the horizon during the radiation era of the
hot Big Bang, because the density ratio $\rho_{\rm GW}/\rho$ is constant, as
the energy density of the gravitational waves (GWs) scales as radiation
\mbox{$\rho_{\rm GW}\propto a^{-4}$} and \mbox{$\rho=\rho_r\propto a^{-4}$} during
the radiation era.  However, as we have seen, during kination
\mbox{$\rho=\rho_{\rm kin}\propto a^{-6}$}, which means that the ratio
\mbox{$\rho_{\rm GW}/\rho$} is not constant any more. In general, this would be
true for any scales which re-enter the horizon during a period with stiff
equation of state, such that \mbox{$\frac13<w\leq 1$}, as
\mbox{$\rho\propto a^{-3(1+w)}$}. 

The density parameter of GWs per logarithmic frequency interval is
\mbox{$\Omega_{\rm GW}(f)\equiv\frac{{\rm d}\Omega_{\rm GW}}{{\rm d}\ln f}$}. Then,
it can be shown that the GW spectrum is of the form \cite{GWspectr}
\begin{equation}
\Omega_{\rm GW}(f)\propto f^{2(\frac{3w-1}{3w+1})},
\label{Omegaf}
\end{equation}
where $w$ is the barotropic parameter at the time when the scales in question
re-enter the horizon. For the scales, which do so during the radiation era where
\mbox{$w=\frac13$} we obtain \mbox{$\Omega_{\rm GW}(f)=\,$constant}, which results
in a scale invariant spectrum (independent of $f$). If there is a period of
kination with \mbox{$w=1$}, then the above suggests
\mbox{$\Omega_{\rm GW}\propto f$} and we have a peak in the spectrum, at the time
when kination begins (usually, the end of inflation; see Fig.~\ref{kin}).
The longer kination is the larger the peak in the GW spectrum. This peak cannot
be arbitrarily large however, because too much GWs can disturb the process of
Big Bang Nucleosynthesis (BBN), as we explain below.

\subsection{Kination and GWs}

\begin{figure}[h]
\vspace{-.3cm}
\centering
\mbox{\hspace{-.5cm}\includegraphics[width=9cm]{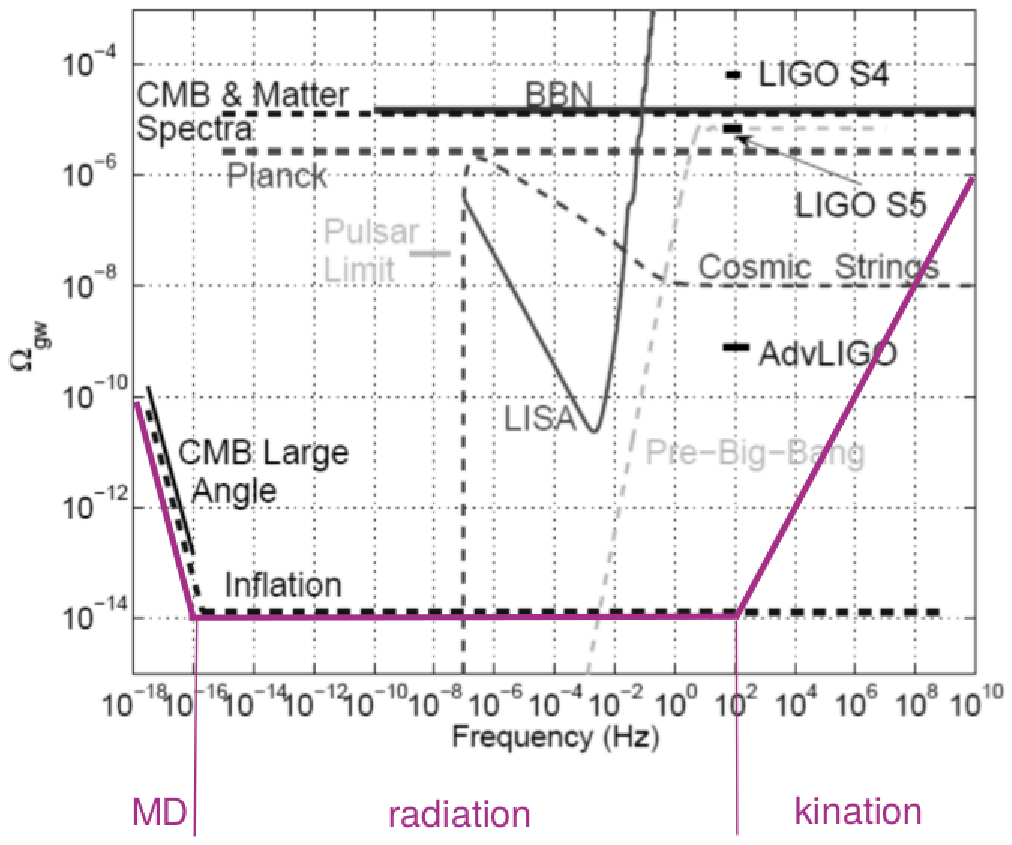}}
\vspace{-5.5cm}
\caption{\label{GWobs1}
  Plot of the GW spectrum per logarithmic frequency interval $\Omega_{\rm GW}$
  with respect to frequency, up to the largest possible frequency which
  corresponds to the inflation energy scale (\mbox{$f_{\rm end}\sim 10^{10}\,$Hz})
  assumed at the scale of grand unification
  \mbox{$\rho_{\rm end}^{1/4}\sim 10^{15-16}\,$GeV}, as is typically the case.
  The solid purple line depicts the spectrum of primordial gravitational waves
  in the case of kination following right after inflation, with barotropic
  parameter \mbox{$w=1$} and reheating temperature
  \mbox{$T_{\rm reh}\sim 10^2\,$MeV}. The peak in the GW spectrum almost saturates
  the BBN bound \mbox{$\Omega_{\rm GW}<10^{-6}$},
  depicted by the horizontal solid black line.
  In the figure, the expected observational capability of
  Advanced LIGO \cite{AdvLIGO} and LISA \cite{LISA}
  are shown. It is evident that this scenario produces unobservable (by LISA and
  Advanced LIGO) primordial GW. Observational bounds have been taken from
  Ref.~\cite{GWobs}.}
\end{figure}

The GW energy density has to be at most 1\% the the time of BBN, i.e.
\mbox{$\Omega_{\rm GW}^{\rm BBN}<10^{-2}$} \cite{GWBBN}.
Now, as we have said, the energy density
of GWs and that of radiation stay at a constant ratio. This means that there is
an upper bound on the GW energy density today, which can be computed as
\begin{eqnarray}
  \Omega_{\rm GW}^0 & = & \frac{\rho_{\rm GW}^0}{\rho^0}\;=\;
  \left.\frac{\rho_{\rm GW}}{\rho_r}\right|_0\left.\frac{\rho_r}{\rho}\right|_0
\nonumber\\
 & = &  \left.\frac{\rho_{\rm GW}}{\rho_r}\right|_{\rm BBN}\Omega_r^0=
  \Omega_{\rm GW}^{\rm BBN}\,\Omega_r^0<10^{-6},
\label{BBNbound}
\end{eqnarray}
where we used that
\mbox{$\Omega_r^0=(\rho_r/\rho)_0\simeq 10^{-4}$} and that
\mbox{$\rho_{\rm GW}/\rho_r=\,$constant}.

This bound means that kination cannot be arbitrary extended to late times (by
considering a small reheating efficiency) because the peak in the GW spectrum
would become too large and violate the bound in Eq.~\eqref{BBNbound}.
Consequently, kination can occur only at very early times, which translates to
high frequencies. A peak in the GW spectrum cannot be extended to lower
frequencies and come in contact with future observations, e.g. of the LISA
space interferometer. This can be understood better in Fig.~\ref{GWobs1}.

\subsection{\boldmath GWs from a stiff period with $w\simeq\frac12$}

One way to avoid violating the bound in Eq.~\eqref{BBNbound} and still extend
the GW peak down to observable frequencies is to consider that, the stiff
period after inflation is not as stiff as kination proper with \mbox{$w=1$}.
Indeed, in Ref.~\cite{figueroa}
it was shown that contact with the LISA observations \cite{LISA}
can be achieved if \mbox{$0.46\lesssim w\lesssim 0.56$} with low reheating
temperature in the range \mbox{1 MeV$\,<T_{\rm reh}\lesssim\,$150 MeV}. A
realisation of this possibility was recently discussed in Ref.~\cite{water}.

\begin{figure}[h]
\centering
  \includegraphics[width=7.5cm]{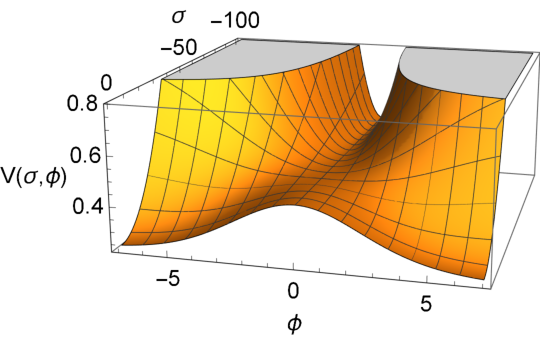}
  \caption{\label{waterfig}
    Qualitative form of the scalar potential in Eq.~\eqref{Vstiff} in terms of
    the canonical scalar fields $\sigma$ (primordial inflaton) and $\phi$
    (waterfall). The axes are in fiducial units. The system origitally finds
    itself gradually rolling down $V(\sigma)$ inside the steep valey at
    \mbox{$\phi=0$} when $\sigma$ is large. When the inflaton is reduced below
    the critical value $\sigma_c$, a phase transition sends the waterfall field
    away from the origin. The minima along the canonical waterfall direction are
    displaced at infinity. As a result, after the phase transition, there is an
    initial period of fast-roll hilltop inflation along the waterfall direction,
    followed by a stiff period. Figure taken by Ref.~\cite{water}.}
\end{figure}

Consider two flat directions meeting at an enhanced symmetry point (ESP).
The scalar potential can be written as
\begin{equation}
V(\varphi,\sigma)=\frac12 g^2\sigma^2\varphi^2+\frac14\lambda(\varphi^2-M^2)^2+
V(\sigma)\,,
\label{Vhybrid}
\end{equation}
where the strength of the interaction
is parametrised by the perturbative coupling constant \mbox{$g<1$}
(we assume that the ESP is at the origin), $\lambda$
is the self-coupling of the $\varphi$-field, whose VEV is $M$ and $V(\sigma)$
is the potential along the $\sigma$-direction. As we discuss below,
\mbox{$M\sim m_P$}, so $\varphi$ is a flat direction lifted by Planck-suppressed
operators. The above potential in Eq.~\eqref{Vhybrid} is the standard
perturbative potential in hybrid inflation \cite{hybrid}, where $\sigma$ plays
the role of the primordial inflaton field. As in standard hybrid inflation,
during
primordial inflation the interaction term sends the waterfall field $\varphi$
to zero, while $V(\sigma)$ provides a gentle slope which allows the inflaton
to slow-roll towards the origin. Primordial inflation is terminated by a phase
transition, when \mbox{$\sigma=\sigma_c\equiv\sqrt\lambda\,M/g$} and the
waterfall field is released from the origin towards its VEV.

\begin{figure}[h]
\vspace{-.3cm}
\centering
\mbox{\hspace{-.5cm}\includegraphics[width=9cm]{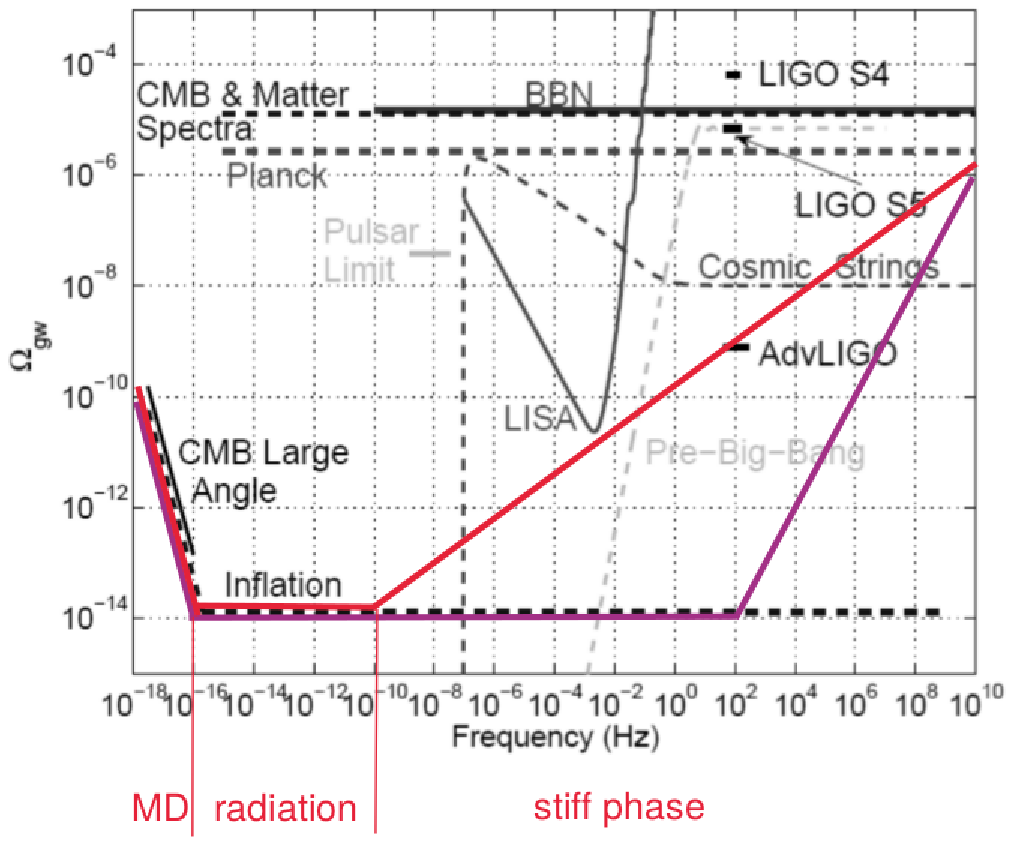}}
\vspace{-5.5cm}
\caption{\label{GWobs2}
  Plot of the GW spectrum per logarithmic frequency interval $\Omega_{\rm GW}$
  with respect to frequency, up to the largest possible frequency which
  corresponds to the inflation energy scale (\mbox{$f_{\rm end}\sim 10^{10}\,$Hz})
  assumed at the scale of grand unification
  \mbox{$\rho_{\rm end}^{1/4}\sim 10^{15-16}\,$GeV}, as is typically the case.
  The solid purple line depicts the spectrum of primordial GWs in the case of
  kination following right after inflation, with barotropic parameter
  \mbox{$w=1$}, while the solid red line depicts the spectrum of primordial GWs
  in the stiff barotropic parameter \mbox{$w=\frac12$} scenario, with reheating
  temperature \mbox{$T_{\rm reh}\sim 10^2\,$MeV}. The peak in the GW spectrum
  almost saturates the BBN bound \mbox{$\Omega_{\rm GW}<10^{-6}$},
  depicted by the horizontal solid black line.
  In the figure, the expected observational capability of Advanced LIGO and LISA
  are shown. It is evident that this scenario produces marginally observable
  (by LISA and Advanced LIGO) primordial GWs. Observational bounds have been
  taken from Ref.~\cite{GWobs}.}
\end{figure}

This is when things become different, because we assume there is a kinetic pole
at the VEV of the waterfall field. Such a pole can be due to some non-trivial
K\"{a}hler geometry as with $\alpha$-attractors \cite{alphattr}.
The Lagrangian density is
\begin{equation}
  {\cal L}=\frac12(\partial\sigma)^2+
  \frac{\frac12(\partial\varphi)^2}{(1-\varphi^2/M^2)^2}-V(\varphi,\sigma)\,.
\label{Lalpha}
\end{equation}
To assist our intuition we redefine the waterfall field so that it also is
characterised by a canonical kinetic term. The redefinition is
\mbox{$\varphi=M\tanh(\phi/M)$}, and $\phi$ is now canonically normalised.
The scalar potential now becomes
\begin{equation}
  V(\phi,\sigma)=\frac12 g^2M^2\sigma^2\tanh^2(\phi/M)+
  \frac{\frac14\lambda M^4}{\cosh^4(\phi/M)}+V(\sigma)\,.
\label{Vstiff}
\end{equation}
The form of the above potential is shown in Fig.~\ref{waterfig}.

After the phase transition which ends primordial inflation, $\sigma$ quickly
goes to zero and the system rolls along the runaway waterfall direction, since
the VEV of $\phi$ is at infinity. Now, exactly because $M$ is Planckian, we have
a further boot of fast-roll inflation \cite{fastroll} as the waterfall field
leaves the origin, which is now a potential hill
\begin{equation}
  V(\phi\ll M)\simeq\frac{\frac14\lambda M^4}{[1+\frac12(\phi/M)^2]^4}\simeq
  \frac14\lambda M^4\left[1-2\left(\frac{\phi}{M}\right)^2\right].
\label{V1}
\end{equation}
Based on the value of $M$, this gives about 13.5 e-folds of hilltop fast-roll
inflation, following the primordial inflation. Inflation ends when
\mbox{$\phi_{\rm end}\simeq M/\sqrt 2$}. Afterwards, the canonical waterfall rolls
along the potential tail with
\begin{equation}
  V(\phi\gg M)\simeq\frac{\frac14\lambda M^4}{[\frac12\exp(\phi/M)]^4}
  \simeq 4\lambda M e^{-4\phi/M}.
\label{V2}
\end{equation}

Since the potential tail is of exponential form, the field soon follows
the power-law inflation attractor \cite{powerlaw} which is characterised by the
barotropic parameter
\mbox{$
w=-1+\frac{16}{3}\left(\frac{m_P}{M}\right)^2.
$}
Requiring \mbox{$w\approx\frac12$} suggests that
\mbox{$M/m_P\approx\frac{4\sqrt 2}{3}=1.88$}. This is why we considered that
\mbox{$M\sim m_P$} in the first place.

As can be seen in Fig.~\ref{GWobs2},
the peak in the GW spectrum is much more mild and
it can come to contact with the future LISA observations.

\subsection{Hyperkination and GWs}

Another possibility to boost the primordial GWs generated during inflation at
observable frequencies is by considering the model of Sec.~\ref{Palatini},
which is characterised by the Lagrangian density in Eq.~\eqref{PalatiniJordan+}.
Switching to Einstein frame through a suitable conformal transformation, the
Lagrangian density becomes
\begin{equation}
{\cal L}=\frac12 m_P^2R+\frac12(\partial\phi)^2+
\alpha\frac{h^2+16\alpha V}{h^2m_P^4}(\partial\phi)^4
-\frac{Vm_P^4}{h^2+16\alpha V}\,,
\label{Lhyp}
\end{equation}  
where \mbox{$h^2\equiv m_P^2+\xi\varphi^2$}, with $\varphi$ being the
non-canonical field, which appears in Eq.~\eqref{PalatiniJordan+}. Then, the
Klein-Gordon equation of motion of the field is
\begin{eqnarray}
  \left[1+12\alpha\left(1+\frac{16\alpha V}{h^2}\right)\right]\ddot\phi\;+
  & & \nonumber\\
  \left[1+4\alpha\left(1+\frac{16\alpha V}{h^2}\right)
    \frac{\dot\phi^2}{m_P^4}\right]3H\dot\phi\;+
  & & \nonumber\\
  48\alpha^2\frac{\dot\phi^4}{m_P^4}
  \frac{\rm d}{\rm d\phi}\left(\frac{V}{h^2}\right)+
\frac{\rm d}{\rm d\phi}\frac{Vm_P^4}{h^2+16\alpha V} & = & 0\,.
\label{KGhyp0}
\end{eqnarray}
From the energy-momentum tensor, it is straightforward to obtain the energy
density and the pressure of the scalar field, which respectively read
\begin{eqnarray}
  \rho_\phi & = & \frac12\left[1+6\alpha\left(1+\frac{16\alpha V}{h^2}\right)
    \frac{\dot\phi^2}{m_P^4}\right]\dot\phi^2+\frac{Vm_P^4}{h^2+16\alpha V}\,,
  \nonumber\\
  & & 
    \label{rhop0}\\
  p_\phi & = & \frac12\left[1+2\alpha\left(1+\frac{16\alpha V}{h^2}\right)
    \frac{\dot\phi^2}{m_P^4}\right]\dot\phi^2-\frac{Vm_P^4}{h^2+16\alpha V}\,.
  \nonumber
\end{eqnarray}

\begin{figure}[h]
\vspace{-.3cm}
\centering
\mbox{\hspace{-.5cm}\includegraphics[width=9cm]{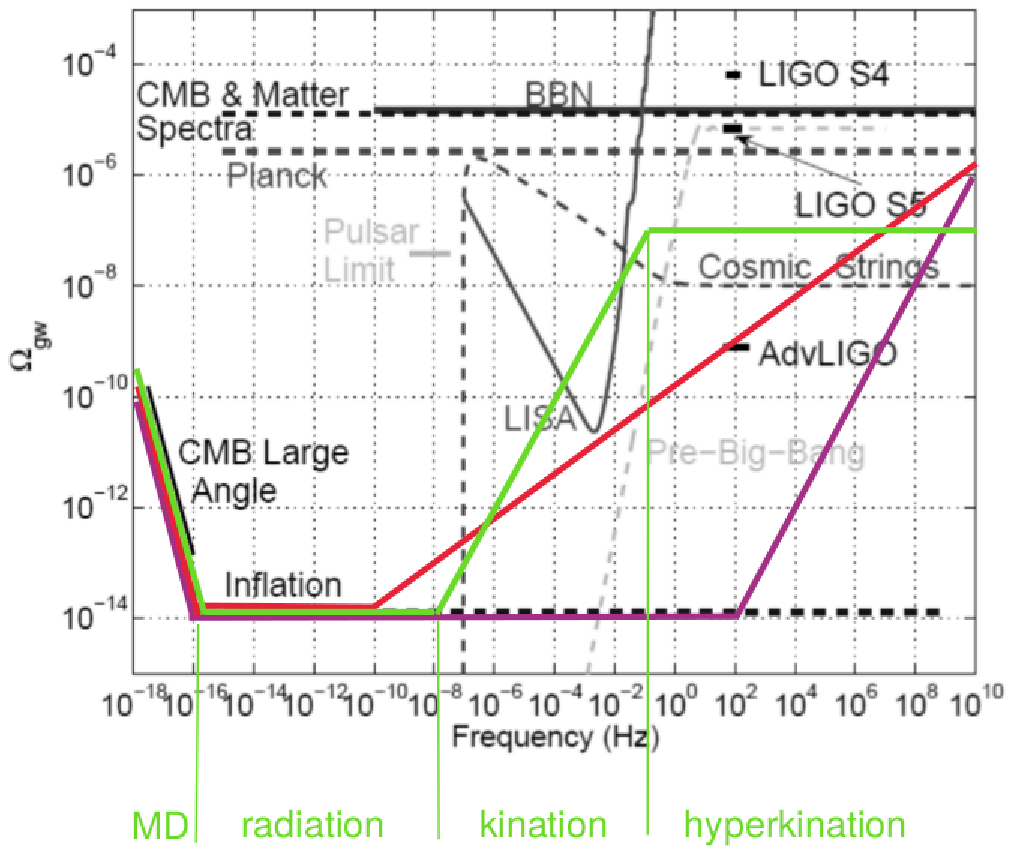}}
\vspace{-5.5cm}
\caption{\label{GWobs3}
  Plot of the GW spectrum per logarithmic frequency interval $\Omega_{\rm GW}$
  with respect to frequency, up to the largest possible frequency which
  corresponds to the inflation energy scale (\mbox{$f_{\rm end}\sim 10^{10}\,$Hz})
  assumed at the scale of grand unification
  \mbox{$\rho_{\rm end}^{1/4}\sim 10^{15-16}\,$GeV}, as is typically the case.
  The solid purple line depicts the spectrum of primordial GWs in the case of
  kination following right after inflation, with barotropic parameter
  \mbox{$w=1$}, while the solid red line depicts the spectrum of primordial GWs
  in the stiff barotropic parameter \mbox{$w=\frac12$} scenario
  and the solid green line depicts the spectrum of primordial GWs in the
  case when inflation is followed first by a period of hyperkination and
  then by a period of regular kination, with reheating temperature
  \mbox{$T_{\rm reh}\sim 10^2\,$MeV}. This time, the spectrum of the GWs is
  nowhere near the BBN bound \mbox{$\Omega_{\rm GW}<10^{-6}$}, depicted by the
  horizontal solid black line. In the figure, the expected observational
  capability of Advanced LIGO and LISA are shown. It is evident that this
  scenario produces a distinct, characteristic spectrum of observable
  (by LISA and Advanced LIGO) primordial GWs. Observational bounds have been
  taken from Ref.~\cite{GWobs}.}
\end{figure}

The above complicated expressions are much simplified if, after the end of
inflation, the scalar field becomes kinetically dominated. In this case, we can
set \mbox{$V\rightarrow 0$} so that Eqs.~\eqref{KGhyp0} and \eqref{rhop0} are
respectively reduced to
\begin{equation}
\left(1+12\alpha\frac{\dot\phi^2}{m_P^4}\right)\ddot\phi+
\left(1+4\alpha\frac{\dot\phi^2}{m_P^4}\right)3H\dot\phi\simeq 0
\label{KGhyp}
\end{equation}  
and
\begin{eqnarray}
  \rho_\phi & \simeq &
  \frac12\left(1+6\alpha\frac{\dot\phi^2}{m_P^4}\right)\dot\phi^2,\nonumber\\
  & & \label{rhop}\\
  p_\phi & \simeq &
  \frac12\left(1+2\alpha\frac{\dot\phi^2}{m_P^4}\right)\dot\phi^2.\nonumber
\end{eqnarray}  

Now, if the standard quadratic kinetic term is dominant in Eq.~\eqref{Lhyp},
then this is equivalent to setting \mbox{$\alpha\rightarrow 0$}. In this case,
Eq.~\eqref{rhop} suggests that \mbox{$w_\phi=p_\phi/\rho_\phi=1$} and we have
regular kination, which results in \mbox{$\Omega_{\rm GW}(f)\propto f$} as we
have discussed. If however, the quartic kinetic term is dominant in
Eq.~\eqref{Lhyp}, then this is equivalent to considering only the terms
proportional to $\alpha$. In this case, Eq.~\eqref{rhop} suggests that
\mbox{$w_\phi=p_\phi/\rho_\phi=\frac13$}, similarly to radiation, for which
\mbox{$\Omega_{\rm GW}(f)=\,$constant}. We call this period
hyperkination~\cite{hyperkin}.

Because, after inflation, the quartic kinetic term
dominates before the quadratic one takes over (and not the other way around),
after inflation we have a period of hyperkination followed by a period of
regular kination until reheating. This results in a GW spectrum, which features
a truncated peak such that boosting the primordial GWs can occur at lower
observable frequencies without violating the BBN bound in Eq.~\eqref{BBNbound}.
The situation is depicted in Fig.~\ref{GWobs3}. The distinct GW spectrum, if
observed, could provide information on the theoretical background, such as the
energy scale of inflation and the value of the $\alpha$ parameter, which
characterises the contribution of the quadratic gravity, as shown in
Eq.~\eqref{PalatiniJordan+}.

\section{Conclusions}

Cosmic Inflation determines the initial conditions of the Universe history and
leads to a large and uniform Universe. Inflation also generates the primordial
density perturbations which seed galaxy formation and are reflected on the
CMB primordial anisotropy. Agreement between the CMB observations and the
predictions of inflation is spectacular. In addition, the Universe today engages
into a late inflationary period, which may be due to quintessence, a form of
dark energy.

Quintessential inflation employs a common theoretical framework for the early
and late Universe and leads to a surge in primordial gravitational waves.
Typically, quintessential inflation is modelled considering a flat direction
with a runaway scalar potential, which has minimum at infinity and features two
flat regions: the inflationary plateau and the quintessential tail. Palatini
gravity is a natural framework for model-building quintessential inflation
because it ``flattens'' a runaway potential to generate the inflationary
plateau.

Quintessential inflation is typically followed by a stiff period called
kination, which generated a peak of primordial gravitational waves (GWs).
However, the kination GW peak corresponds to unobservable frequencies. One way
to overcome this is by considering that the peak is milder, which can be
achieved in the stiff period after inflation is not as stiff as kination proper.
A model realisation of this possibility considers two flat directions
intersecting at an enhanced symmetry point in field space, giving rise to the
hybrid mechanism, with Planckian waterfall field VEV, which is also a kinetic
pole of the waterfall field, following the $\alpha$-attractors proposal.

Another interesting possibility to obtain a boost of primordial GWs down at
observable frequencies in by considering higher order kinetic terms (as in
k-essence \cite{kess}).
This is possible to realise in Palatini modified gravity. Indeed,
considering $R+R^2$ gravity and a non-minimally coupled scalar field gives rise
to additional quartic kinetic terms. When these dominate, this leads to
hyperkination which is followed by regular kination when the field becomes
canonical. the resulting characteristic truncated GW peak can be extended to
observable frequencies without disturbing BBN.

Forthcoming observations of Advanced LIGO, LISA, DesiGO or BBO may well detect
the primordial GWs generated by inflation. A distinct GW spectrum will provide
insights to the background theory. Finally, it should be pointed out that the
detection of primordial GWs will not only confirm another prediction of cosmic
inflation but also offer tantalising evidence for the quantum nature of gravity
itself.

\end{document}